\documentclass[showpacs,prb,aps,superscriptaddress,twocolumn,floatfix]{revtex4}
\usepackage{graphicx,float,amsmath,amssymb}

\newfont{\ensmathquatorze}{msbm10 scaled 1400}
\newfont{\ensmathonze}{msbm10 scaled 1100}
\newfont{\ensmathdix}{msbm10}
\newfont{\ensmathneuf}{msbm10 scaled 833}
\newfont{\ensmathhuit}{msbm10 scaled 694}
\newfam\ensmathfam                        
\textfont\ensmathfam=\ensmathonze        
\scriptfont\ensmathfam=\ensmathdix       
\scriptscriptfont\ensmathfam=\ensmathhuit

\renewcommand{\leq}{\leqslant}

\newcommand{\ket}[1]{|\kern.3ex#1\kern.3ex\rangle}
\newcommand{\bra}[1]{\langle\kern.3ex #1 \kern.3ex|}

\newcommand{\smean}[1]{\langle #1 \rangle} 

\newcommand{\EXP}[1]{{\mbox{\large e}}^{#1}}         

\def\I{{\rm i}}                  
\def\D{{\rm d}}                  

\newcommand{\diagram}[3]{\raisebox{#3}{\includegraphics[scale=#2]{#1}}}

\begin{document}

\title{Effect of connecting wires on the decoherence due to\\ electron-electron interaction in a metallic ring}

\author{Christophe Texier}
\affiliation{Laboratoire de Physique Th\'eorique et Mod\`eles Statistiques,
              UMR 8626 du CNRS, Universit\'e Paris-Sud, 91405 Orsay, 
              France.}
\affiliation{Laboratoire de Physique des Solides, UMR 8502 du CNRS,
              Universit\'e Paris-Sud, 91405 Orsay, France.}

\date{July 19, 2007}

\begin{abstract}
  We consider the weak localization in a ring connected to reservoirs through
  leads of finite length and submitted to a magnetic field. The effect of
  decoherence due to electron-electron interaction on the harmonics of AAS
  oscillations is studied, and more specifically the effect of the leads. Two
  results are obtained for short and long leads regimes. The scale at which
  the crossover occurs is discussed. The long leads regime is shown to be more
  realistic experimentally.
\end{abstract}

\pacs{73.23.-b~; 73.20.Fz~; 72.15.Rn}

\maketitle


The classical Drude conductivity of metals is corrected at low temperature by
small quantum corrections. One of them, called the {\it weak localization (WL)
  correction}, is a contribution to the average conductance due to
interferences of reversed electronic trajectories. WL is a manifestation of
quantum coherence, which provides a powerful tool to study phase coherence in
weakly disordered metals. A possible way to measure WL is to analyze its
magnetic field dependence~: in particular the conductivity of a ring presents
$\phi_0/2$-periodic oscillations, as a function of the magnetic flux piercing
the ring, called the Al'tshuler-Aronov-Spivak (AAS)
oscillations~\cite{AltAroSpi81,ShaSha81} ($\phi_0=h/e$ is the quantum flux).
Measurement of AAS harmonics provides a central quantity of mesoscopic
physics~: the {\it phase coherence length}, the typical length over which
electronic phase coherence is ensured.

Two articles~\cite{LudMir04,TexMon05b} 
have recently discussed how
AAS harmonics of the WL correction to conductance $\smean{\Delta{g}_n}$ (or
Aharonov-Bohm (AB) harmonics) of a metallic ring are affected by the
decoherence due to electron-electron interaction. Ref.~\cite{LudMir04}
emphasized the role to the wires connecting the ring to reservoirs, contrary
to Ref.~\cite{TexMon05b}. This is the aim of the present paper to clarify this
question.

We recall the basic framework necessary to study the question. We consider a
metallic ring of perimeter $L$, connected at two reservoirs through leads of
lengths $l_a$ (we consider a symmetric situation where the two arms of the
ring have equal lengths $L/2$). We define the geometrical
parameter~\cite{LudMir04} $\gamma=(8l_a/L+1)^{-1}$. AAS harmonics involve an
integral of the cooperon $P_c^{(n)}(x,x)$ inside the network where wires are
appropriately weighted~\cite{TexMon04,TexMon05b}~:
\begin{equation}
  \label{harmWL}
  \smean{\Delta{g}_n} \simeq 
  -8\,\gamma^2
  \int_\mathrm{ring}\frac{\D x}{L^2}\,P_c^{(n)}(x,x)
  \:.
\end{equation}
$P_c^{(n)}(x,x)$ encodes the contribution of closed interfering reversed
electronic trajectories issuing from $x$ with winding $n$. In the present
article we will be only interested in the limit of strong decoherence (high
temperature)~\cite{footnote1}, therefore, since the harmonics of the cooperon
are expected to vanish exponentially inside the leads on a scale given by the
phase coherence length, contributions $\int_\mathrm{leads}\D{x}\,P_c^{(n)}$ of
the connecting wires have been neglected in (\ref{harmWL}). Decoherence due to
electron-electron interaction is taken into account within the framework of
Ref.~\cite{AltAroKhm82}~: its efficiency is characterized by the Nyquist
length $L_N=({\nu_0D^2}/{T})^{1/3}$ where $\nu_0$ is the density of
states, 
$D$ the diffusion constant and $T$ the temperature. The cooperon can be
written as a path integral
\begin{eqnarray}
  \label{cooperon}
  P_c^{(n)}(x,x)
  =\int_0^\infty\D{t}
  \int_{x(0)=x}^{x(t)=x}\hspace{-0.5cm}\mathcal{D}x(t')\,
  \delta_{n,\mathcal{N}[x(t')]}\,\nonumber\\
  \times
  \EXP{ -\int_0^t\D{t'}\,\big[\frac14\dot{x}^2+\frac{2}{L_N^3}W(x(t'),x(t-t'))\big] }
  \:,
\end{eqnarray}
where $\mathcal{N}[x]$ is winding number of a diffusive trajectory around the
ring. In principle path integral runs over trajectories $x(t')$ that explore
the whole network, however in the limit of strong decoherence ($L_N\ll{L}$) we
will make the approximation that they remain inside the ring. Decoherence due
to electron-electron interaction is modeled as phase fluctuations for a given
electron due to coupling to the fluctuating electromagnetic field created by
other electrons~\cite{AltAroKhm82,AltAro85,AkkMon07}. Decoherence is accounted
in (\ref{cooperon}) through the functional
$\exp{-\frac{2}{L_N^3}\int_0^t\D{t'}W}$, where $W$ is related to the
correlation function of electric potential (given by the
fluctuation-dissipation theorem~\cite{AltAro85,AkkMon07}). Therefore, decoherence
depends on the network, and moreover on the nature of trajectories. In
(\ref{harmWL}) the leads seem absent at first sight, however their effect is
hidden in the cooperon, since the leads affect the nature of fluctuations of
the electric potential, and therefore the decoherence, through $W$. It is the
purpose of the present paper to discuss if it is an important effect or not.
We will study successively the long and short lead limits. Finally we will
obtain the scale of the crossover between the two regimes and their
experimental relevance.

\vspace{0.15cm}

\mathversion{bold}
\noindent{\bf Long leads~: $\gamma\simeq0$.--}
\mathversion{normal} The difficulty in evaluating
(\ref{harmWL},\ref{cooperon}) lies in the time nonlocality of the action. In
the case $\gamma=0$, translation invariance inside the ring is recovered,
$W(x,x')=W(x-x',0)$, what allows to get rid of time nonlocality thanks
to~\cite{TexMon05b} $W(x(t')-x(t-t'),0)\to{W}(x(t'),0)$, and compute precisely
the path integral. The cooperon is equivalent to the Green function describing
tunneling of a particle with zero energy through periodic series of quadratic
barriers, $V(x)=\frac{4}{L_N^3}W(x,0)\propto{x}(1-\frac{x}{L})$ for
$x\in[0,L]$, starting from $x=0$ and arriving at $x=nL$. For~$L_N\ll{L}$, we
have found in Ref.~\cite{TexMon05b}~:
\begin{equation}
  P_c^{(n)}(x,x)\sim L_N\,\EXP{-nS_\mathrm{inst}}
  \ \ \forall\: x\in\mbox{ ring}
  \:.
\end{equation}
$S_\mathrm{inst}=\frac\pi8(\frac{L}{L_N})^{3/2}$ is the action of the
tunneling trajectory through one barrier. The prefactor $L_N$ comes from the
linear behaviour of the potential $W(x,x')\simeq\frac12|x-x'|$ at short
distance, corresponding to $V(x)\propto|x|$ at the initial and final point for
the tunneling trajectory~\cite{footnoteBadLM}. Finally the AAS harmonics of WL
correction to the conductance read~:
\begin{equation}
  \label{resultAAS}
  \smean{\Delta g_n} \sim -\gamma^2\,
  \frac{L_N}{L}\, \EXP{ -n\frac\pi8(\frac{L}{L_N})^{3/2} }
  \:.
\end{equation}
Harmonic of AB oscillations are related to AAS harmonics by~\cite{footnote3}~:
$\smean{\delta{g}^2_n}\sim-\gamma^2(\frac{L_T}{L})^2\smean{\Delta{g}_n}$,
where $L_T=\sqrt{D/T}$ is the thermal length, which leads to the temperature
dependence~$\smean{\delta{g}^2_n}\propto(\frac{L_T}{L})^{8/3}\EXP{-nS_\mathrm{inst}}$.
Two possible interpretations can be given to the result
$\smean{\Delta{g}_n}\propto\smean{\delta{g}^2_n}\propto\EXP{-nS_\mathrm{inst}}$.
The scaling of harmonics with perimeter $L$ suggests to define a phase
coherence length as $L_\varphi\to{L}_N\propto{T}^{-1/3}$. On the other hand,
assuming the AAS~\cite{AltAroSpi81} form
$\smean{\Delta{g}_n}\propto\EXP{-nL/L_\varphi}$, that is to consider the
scaling with $n$, leads to define a $L$-dependent phase coherence length
$L_\varphi\to\frac8\pi\frac{L_N^{3/2}}{L^{1/2}}\propto{T}^{-1/2}$. Physical
interpretation of this new length scale have been given in
Refs.~\cite{LudMir04,TexMon05b}.

\vspace{0.15cm}

\mathversion{bold}
\noindent{\bf Short leads~: $\gamma\neq0$.--}
\mathversion{normal} In this case $W$ is not anymore translation invariant
inside the ring. To circumvent the problem of time nonlocality of the action,
we can use the following procedure~: for $t'\in[0,t/2]$ we introduce
$x_1(t')=x(t')$ and $x_2(t')=x(t-t')$. Thanks to the property
$W(x,x')=W(x',x)$ we get~:
\begin{eqnarray}
  \label{cooperon2}
  P_c^{(n)}(x,x)
  =2\int_0^\infty\D{t}\int_\mathrm{ring}\hspace{-0.25cm}\D{x'}
  \int_{x_1(0)=x}^{x_1(t)=x'}\hspace{-0.85cm}\mathcal{D}x_1(t')\,
  \int_{x_2(0)=x}^{x_2(t)=x'}\hspace{-0.85cm}\mathcal{D}x_2(t')\,
  \nonumber\\
  \times\delta_{n,\mathcal{N}[x_1,x_2]}\,
  \EXP{ -S[x_1,x_2] }
  \:,\hspace{1cm}
\end{eqnarray}
where $\mathcal{N}[x_1,x_2]$ is the winding number of the two trajectories put
end to end. The action 
$
S[x_1,x_2] = \int_0^t\D{t'}\,
\big[\frac14(\dot{x}_1^2+\dot{x}_2^2) + \frac{4}{L_N^3}W(x_1(t'),x_2(t'))\big] 
$
describes a 2-dimensional problem, local in time. In the high
temperature regime of interest here ($L_N\ll{L}$) we can use semiclassical
methods. Path integral (\ref{cooperon2}) is dominated by classical
trajectories of winding $n$. Such trajectories are solutions of the classical
equation of motion $\ddot{\vec{r}}=\frac8{L_N^3}\vec\nabla{W}$ on the torus
$[0,L]\times[0,L]$, where $\vec{r}\equiv(x_1,x_2)$, for initial and final
conditions $\vec{r}(0)=(x,x)$ and $\vec{r}(t)=(x',x')$, with zero energy
[hamiltonian is
$\mathcal{H}=-\frac14\dot{\vec{r}}\,^2+\frac4{L_N^3}W(\vec{r})$]. To simplify
the problem we can set $L=L_N=1$, and denote by $\widetilde{S}$ the new
dimensionless action, since at zero energy the $L/L_N$ dependence of the
action is easily recovered~:
$S_{\mathcal{H}=0}=(\frac{L}{L_N})^{3/2}\widetilde{S}_{\mathcal{H}=0}$. At
zero energy, the initial condition on the line $x_1=x_2$, such that $W=0$,
fully determines the classical trajectory. Therefore we expect the structure
$P_c^{(n)}(x,x)\propto\EXP{-n(\frac{L}{L_N})^{3/2}\widetilde{S}^\mathrm{cl}(x/L)}$,
where $\widetilde{S}^\mathrm{cl}(x)$ is the action of the classical trajectory
issuing from $\vec{r}(0)=(x,x)$ with winding $n=1$ (for $L=L_N=1$).

\begin{figure}[!ht]
  \centering
  \diagram{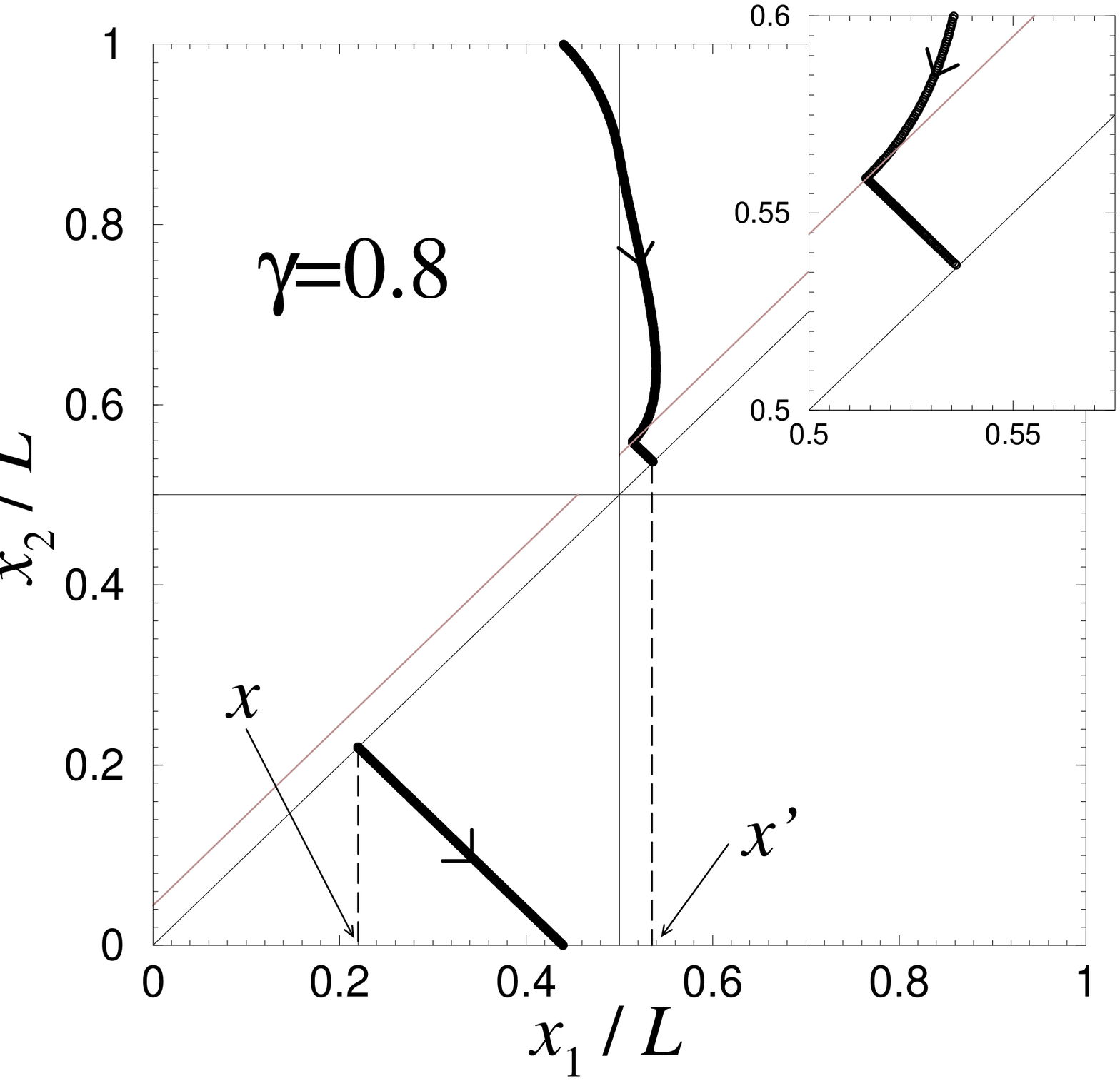}{0.25}{0cm}
  \hspace{0.5cm}
  \diagram{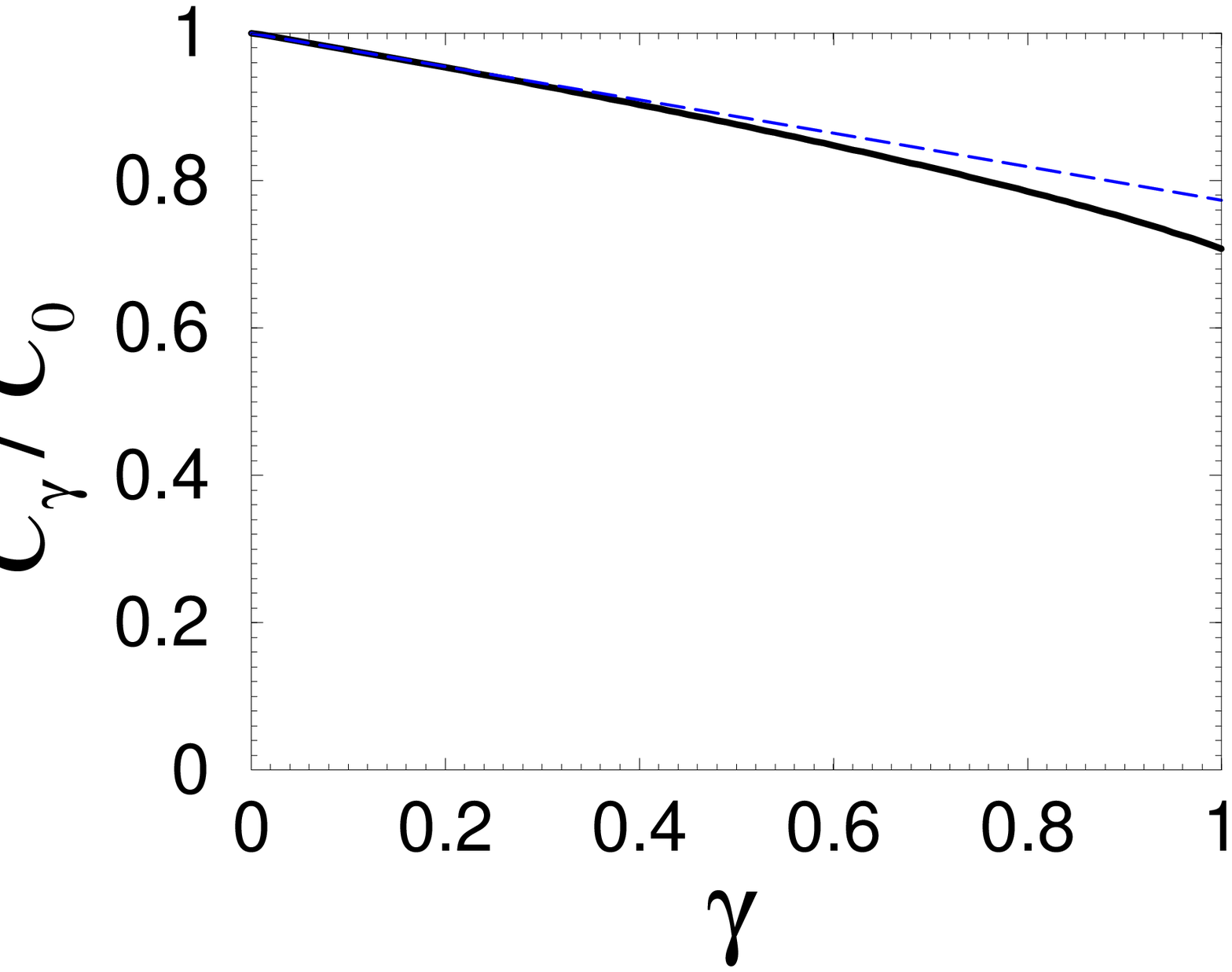}{0.2}{0.5cm}
  \caption{
    Left~~:
      {\it Example of a classical trajectory of winding $n=1$. 
      The last small segment
      (inset) corresponds to tunneling~: coordinates aquire a complex part not
      represented.}      
          }
  \label{fig:CgamTraj}
\end{figure}

\begin{figure}[!ht]
  \centering
  \diagram{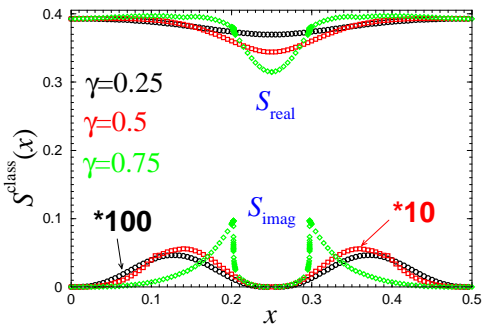}{0.375}{0cm}
 
  \vspace{-0.6cm}

  \caption{\it Real (top curves) and imaginary (bottom curves) parts of the
    action of classical trajectory as a function of initial condition $x$, for
    $\gamma=0.25$ (black circles), $\gamma=0.5$ (red squares) and
    $\gamma=0.75$ (green diamonds). Imaginary parts for $\gamma=0.25$ and
    $\gamma=0.5$ have been rescaled by factors $100$ and $10$, respectively.}
  \label{fig:action}
\end{figure}

Thanks to the symmetries of potential, we can easily find two particular
solutions of the equation of motion.
({\it i})~The solution on the line $x_1+x_2=L/2$ (or $3L/2$) is the one with
minimal action
$\widetilde{S}^\mathrm{cl}(1/4)=\frac{C_\gamma}{C_0}\,\frac\pi8$. The
dimensionless parameter is~\cite{LudMir04}~:
$C_\gamma=(\frac{\pi}{2(\gamma+1)})^{3/2}[\pi+2\arcsin\gamma+2\gamma\sqrt{1-\gamma^2}]$. 
The ratio
$C_\gamma/C_0$ varies monotonously between $1$ and $C_1/C_0=1/\sqrt2$ 
(Fig.~\ref{fig:CgamTraj}).
({\it ii})~The solution on the line $x_1+x_2=0$ (or $L$) is the one
with maximal action $\widetilde{S}^\mathrm{cl}(0)=\frac\pi8$.
Apart for initial conditions $x=0$, $L/4$, $L/2$ or $3L/4$, the trajectory
aquires some finite speed $v_\parallel=\frac{\dot{x}_1+\dot{x}_2}{\sqrt2}$ in
the direction parallel to the line $x_1=x_2$. Therefore it can only wind
around the ring by going to complex time $t'=\pm\I\tau$. Action also aquires
some complex part,
$\widetilde{S}^\mathrm{cl}(x)=S_\mathrm{real}(x)+\I{S}_\mathrm{imag}(x)$, and
can be studied numerically (Fig.~\ref{fig:action}). (There are two
complex trajectories for the two possible signs $t'=\pm\I\tau$, with conjugated
actions).

Finally, assuming the structure
\begin{equation}
  P_c^{(n)}(x,x)\sim L_N\,\EXP{-n(\frac{L}{L_N})^{3/2}\widetilde{S}^\mathrm{cl}(x/L)}
  \:,
\end{equation}
where we have kept the same preexponential factor $L_N$ as for the case
$\gamma=0$ since potential presents the same short distance behaviour
$W\simeq\frac12|x-x'|$. The cooperon is maximal at the middle of the arms of
the ring ($x=L/4$ and $3L/4$). Assuming
Gaussian fluctuations and using the steepest descent approximation, we
obtain~:
\begin{equation}
   \label{resultAAS2}
   \smean{\Delta g_n} \sim -\frac{\gamma^2}{\sqrt{n}}
   \left(\frac{L_N}{L}\right)^{7/4}\,
   \EXP{-n \frac{C_\gamma}{C_0}\,\frac\pi8(\frac{L}{L_N})^{3/2} } 
  \:.
\end{equation}
Now we write $S_\mathrm{inst}=\frac{C_\gamma}{C_0}\,\frac\pi8(\frac{L}{L_N})^{3/2}$.
Integration over $x$ has brought an additional factor $(L_N/L)^{3/4}$.
Corresponding AB harmonics are
$\smean{\delta{g}_n^2}\propto(\frac{L_T}{L})^{19/6}\EXP{-nS_\mathrm{inst}}$.
Despite we follow the same ideas as LM,\cite{LudMir04} our preexponential
factors for AB/AAS amplitudes disagree~\cite{footnoteBadLM} with LM's results
$\smean{\Delta{g}_n}_\mathrm{LM}\sim{L}_N^{9/4}\EXP{-nS_\mathrm{inst}}$ and
$\smean{\delta{g}_n^2}_\mathrm{LM}\sim{L}_T^{7/2}\EXP{-nS_\mathrm{inst}}$, by
a factor~$(L_N/L)^{1/2}\propto{T}^{-1/6}$.

\vspace{0.15cm}

\noindent{\bf Crossover.--}
The steepest descent method, used to derive (\ref{resultAAS2}), requires
$(1-\frac{C_\gamma}{C_0})\frac\pi8(\frac{L}{L_N})^{3/2}\gg1$. In the opposite
limit, $P_c^{(n)}(x,x)$ can be considered almost constant inside the ring and
eq.~(\ref{resultAAS}) holds. Therefore, since
$\frac{C_\gamma}{C_0}\simeq1-0.23\gamma$ (where
$\frac32-\frac4\pi\simeq0.23$), the crossover between these two situations
occurs for $\gamma\sim\gamma_c$ where
\begin{equation}
  \gamma_c \simeq 11\left(\frac{L_N}{L}\right)^{3/2}
  \:.
\end{equation}

\vspace{0.15cm}

\noindent{\bf Discussion.--}
To summarize, we have obtained two expressions (\ref{resultAAS}) and
(\ref{resultAAS2}) for the AAS harmonics, corresponding to long
($\gamma\lesssim\gamma_c$) and short leads ($\gamma\gtrsim\gamma_c$) regime.
Their derivations assumed~:
({\it i})~$L_N/L\ll1$ (validity of semiclassical approximation),
({\it ii})~$L_N/l_a\ll1$, where $l_a$ is the length of the leads, otherwise some
other $L_N$-dependent preexponential factor is
expected~\cite{TexMon05,TexMon05b} since winding trajectories feel absorbing reservoirs.
From experimental point of view, observability of the effect requires
({\it iii})~$L_N/L$ large enough in order to obtain some signal.
Applicability of long leads result (\ref{resultAAS}) is
({\it iv})~$l_a/L$ large ($\gamma\lesssim\gamma_c$).
On the other hand, applicability of the short leads result (\ref{resultAAS2})
requires~:
({\it v})~$l_a/L$ small enough
$\frac{l_a}{L}\lesssim\frac18(\frac1{\gamma_c}-1)$, what is more problematic.
Let us rewrite contraints ({\it ii}) \& ({\it v}) as
$\frac{l_a}{L_N}\gtrsim1$ and
$\frac{l_a}{L_N}\lesssim\frac18\frac{L}{L_N}[0.09(\frac{L}{L_N})^{3/2}-1]$.
This is only possible for $L_N/L\lesssim1/8$, which seems pretty irrealistic
from experimental point of view since it corresponds to a suppression of AAS
harmonics of~$\EXP{-\frac\pi8(\frac{L}{L_N})^{3/2}}\approx10^{-4}$. 

Let us consider a real situation and use parameters corresponding to
experiments of Ref.~\cite{FerAngRowGueBouTexMonMai04}. The highest temperature
at which AAS oscillations were still
observable~\cite{FerAngRowGueBouTexMonMai07} was $T=700\:$mK (the Nyquist
length was $L_N\simeq0.59\:\mu$m at 1\:K) at which $L_N/L\simeq0.17$, which is
above the threshold of applicability of the short leads result (\ref{resultAAS2}).
At this temperature $\gamma_c\simeq0.7$, which leaves a large room to the long
leads regime (\ref{resultAAS}), and for which ({\it ii}) is not possible to
fulfill since ({\it v}) gives $l_a\lesssim0.2\:\mu$m far
below~$L_N=0.66\:\mu$m.
However we note that, up to now, the  experiment~\cite{FerAngRowGueBouTexMonMai07}
is not able to give information on the $T$-dependent prefactor.

We conclude that the long leads result (\ref{resultAAS}) might be used for devices
easily realized, whereas short leads result (\ref{resultAAS2}) seems to apply
to rather irrealistic situations.

\vspace{0.15cm}

\noindent{\bf Acknowlegments.--}
I thank Gilles~Montambaux and Denis~Ullmo for interesting discussions and comments.

\vspace{0.15cm}

\mathversion{bold}
\noindent{\bf Appendix~: potential $W(x,x')$.--}
\mathversion{normal} 
The potential is defined~\cite{TexMon05b} as
$W(x,x')=\frac12[P_d(x,x)+P_d(x',x')]-P_d(x,x')$, where the
diffuson $P_d$ is solution of $-\Delta{P}_d(x,x')=\delta(x-x')$ and vanishes
at the reservoirs.
By definition it is a symmetric function of its two arguments.
It has fully been constructed for a connected ring in Ref.~\cite{TexMon05b}.
We choose coordinate $x\in[0,L]$ such that $x=0$ and $x=L/2$ correspond to the
points of attachment of the leads.

\noindent$\bullet$ If $x,\,x'\in$ the same arm
$W(x,x')=\frac12|x-x'|[1-(\gamma+1)\frac{|x-x'|}{L}]$.
In the plane $(x,x')$, equipotentials are straight lines.

\noindent$\bullet$ If $x,\,x'\notin$ the same arm.
For $0\leq x\leq{L/2}\leq{x'}\leq{L}$~:
$
W(x,x') = \frac12(x'-x)[1-(\gamma+1)\frac{x'-x}{L}]
-\frac{2\gamma}{L}(x-\frac{L}2)(x'-\frac{L}2)
$.
Equipotentials are ellipses (circles for $\gamma=1$).
Note that on the line $x+x'=L$ we find that
$W(x,L-x)=x(1-\frac{2x}{L})$ is independent on~$\gamma$.

\begin{figure}[!ht]
  \centering
  \diagram{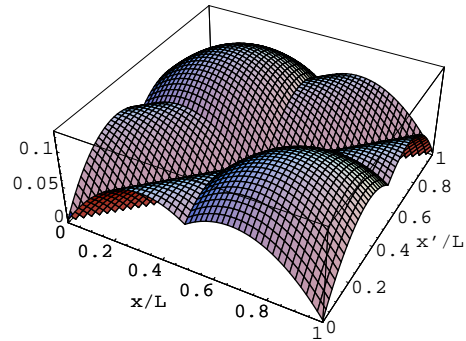}{0.75}{0cm}
 
  \vspace{-0.5cm}

  \caption{\it Potential $W(x,x')$ as a function of $x/L$ and $x'/L$, 
               for $\gamma=0.5$.}
  \label{fig:fctW}
\end{figure}

\vfill


\begin{thebibliography}{20}
\bibitem{footnote1}
  The high temperature regime (phase coherence length larger than perimeter)
  has been studied in Ref.~\cite{TexMon05b}.
\bibitem{footnoteBadLM}
  In Ref.~\cite{LudMir04}, LM have unproperly used semiclassical methods to
  estimate preexponential factor. In particular they did not account for the
  fact that initial/final point in path integral are turning points where WKB
  approximation fails.
\bibitem{footnote3} As noticed in appendix E of Ref.~\cite{TexMon05b}, the
  relation between AB and AAS harmonics is not universal but holds when the
  role of the wire weights introduced in Ref.~\cite{TexMon04} can be
  neglected. For the symmetric connected ring considered here, we distinguish
  two limits. 
  (1) $L_\varphi\gg{L}$~: quantum correction is dominated by integrals over
  the leads and the relation
  $
  \smean{\delta{g}^2_n}\simeq
  -\frac{16\pi}{3}\gamma^2(\frac{L_T}{L})^2\smean{\Delta{g}_n}
  \simeq-\frac{\pi}{3}(\frac{L_T}{2l_a})^2\smean{\Delta{g}_n}
  $ 
  is exactly similar as for a long wire~\cite{AleBla02}.
  (2) $L_\varphi\ll{L}$~: quantum correction is dominated by integrals inside 
  the ring, whose arms have a relative weight $1/4$ compare to the leads~: 
  $
  \smean{\delta{g}^2_n}\simeq
  -\frac{4\pi}{3}\gamma^2(\frac{L_T}{L})^2\smean{\Delta{g}_n}
  $.
  These relations are valid for $L_\varphi\gg{L_T}$, which is always the case 
  if electron-electron is the dominant mechanism~\cite{AltAro85} described by
  the phase coherent length~$L_\varphi$.
\end{thebibliography}

\end{document}